\documentclass[12pt]{article}

\begin{document}

\begin{flushright}
SLAC-PUB-9107\\
January 2002
\end{flushright}

\bigskip\bigskip

\begin{center}
{{\bf\Large Hadron Spin Dynamics\footnote{Work supported by the
Department of Energy, contract DE--AC03--76SF00515.}}}

\vfill
Stanley J. Brodsky \\
Stanford Linear Accelerator Center, Stanford University \\
Stanford, California 94309\\
e-mail: sjbth@slac.stanford.edu

\vfill

\begin{abstract}

Spin effects in exclusive and inclusive reactions provide an
essential new dimension for testing QCD and unraveling hadron
structure. Remarkable new experiments from SLAC, HERMES (DESY),
and Jefferson Lab present many challenges to theory, including
measurements at HERMES and SMC of the single spin asymmetries in
$e p \to e' \pi X$ where the proton is polarized normal to the
scattering plane.  This type of single spin asymmetry may be due
to the effects of rescattering of the outgoing quark on the
spectators of the target proton, an effect usually neglected in
conventional QCD analyses. Many aspects of spin, such as
single-spin asymmetries and baryon magnetic moments are sensitive
to the dynamics of hadrons at the amplitude level, rather than
probability distributions.  I will illustrate the novel features
of spin dynamics for relativistic systems by examining the
explicit form of the light-front wavefunctions for the
two-particle Fock state of the electron in QED, thus connecting
the Schwinger anomalous magnetic moment to the spin and orbital
momentum carried by its Fock state constituents and providing a
transparent basis for understanding the structure of relativistic
composite systems and their matrix elements in hadronic physics.
I also present a survey of outstanding spin puzzles in QCD,
particularly $A_{NN}$ in elastic $pp$ scattering, the $J/\psi \to
\rho \pi$ puzzle, and $J/\psi$ polarization at the Tevatron.

\end{abstract}

Concluding Theory Talk, presented at the \\
3rd Circum-Pan-Pacific Symposium On High Energy Spin Physics (SPIN
2001)\\
Beijing, China \\
8--13 October 2001
\end{center}
\vfill

\newpage

\newcommand{\ie}{{\it i.e.}}
\newcommand{\eg}{{\it e.g.,}}
\newcommand{\btt}[1]{{\tt$\backslash$#1}}
\newcommand{\half}{{$\frac{1}{2}$}} 
\newcommand{\ket}[1]{\left\vert\,{#1}\right\rangle}
\newcommand{\VEV}[1]{\left\langle{#1}\right\rangle}
\renewcommand{\bar}[1]{\overline{#1}}
\newcommand{\qu}{{\rm q}}
\newcommand{\qb}{${\rm\bar q}$}
\newcommand{\qbm}{{\rm\bar q}}
\newcommand{\pvec}{\vec p}
\newcommand{\kvec}{\vec k}
\newcommand{\rvec}{\vec r}
\newcommand{\Rvec}{\vec R}
\newcommand{\lqcd}{\Lambda_{QCD}}
\newcommand{\ieps}{i\varepsilon}
\newcommand{\disc}{{\rm Disc}}
\newcommand{\pl}{{||}}
\newcommand{\order}[1]{${ O}\left(#1 \right)$}
\newcommand{\morder}[1]{{ O}\left(#1 \right)}
\newcommand{\eq}[1]{(\ref{#1})}
\newcommand{\beq}{\begin{equation}}
\newcommand{\eeq}{\end{equation}}
\newcommand{\beqa}{\begin{eqnarray}}
\newcommand{\eeqa}{\end{eqnarray}}
\newcommand{\etal}{{\em et al.}}
\def\ru1{\rule[-0.4truecm]{0mm}{1truecm}}
\def\upleftarrow#1{\overleftarrow{#1}}
\def\uprightarrow#1{\overrightarrow{#1}}
\def\thru#1{\mathrel{\mathop{#1\!\!\!/}}}

\section{Introduction}

One of the most important goals in quantum chromodynamics is to
determine the fundamental structure of hadrons in terms of their
quark and gluon degrees of freedom.  Spin-dependent phenomena play
a crucial role in this pursuit, dramatically increasing the
sensitivity to important and subtle QCD
effects~\cite{Filippone:2001ux}.  Many aspects of spin
phenomenology,  such as the proton's magnetic moment and
single-spin asymmetries, require an understanding of hadron
structure at the amplitude level, rather than probability
distributions.  Spin phenomena include such topics as the
magnetic moments of baryons, nuclear polarization effects in
deuteron photo-disintegration,  spin correlations in $B$ decays,
spin-spin correlations in high momentum transfer exclusive and
inclusive reactions, and the spin correlations of top quarks
produced in high energy colliders.  The upcoming RHIC spin program
will study high energy polarized proton-proton collisions, with
the potential of directly measuring the spin carried by gluons in
the proton~\cite{Bunce:2000uv,Bass:2001st,Ma:2001na}.  The planned
charm photoproduction studies at SLAC with polarized photons and
protons will provide important checks on leading-twist QCD
predictions~\cite{Bosted:pd}.  New programs at
HERMES~\cite{Heinsius:2001eg} and Jefferson
Laboratory~\cite{Burkert:2001cc} will study spin structure
functions, spin correlations, and azimuthal polarization
asymmetries in deeply virtual Compton scattering and in other
exclusive channels.

A number of new experimental results from SLAC, HERMES (DESY),
COMPASS at CERN, and Jefferson Lab were reported at this
meeting~\cite{Peng}, presenting many new challenges to theory.
Most hadron spin experiments have been performed with protons;
however there are also potentially important tests of QCD
involving higher spin targets such as the
deuteron~\cite{Hoodbhoy:1990yv,Brodsky:1992px,Brodsky:2001qm}
which test nuclear coherence and concepts such as hidden
color~\cite{Brodsky:1983vf}.  Some of the most interesting
confrontations with theory have arisen from the measurements by
the HERMES \cite{hermes0001} and SMC \cite{smc99} collaborations
of single-spin asymmetry in inclusive electroproduction $e p \to
e' \pi X$ where the proton is polarized normal to the
photon-to-pion scattering plane.  In the case of HERMES, the proton
is polarized along the incident electron beam direction, and thus
the azimuthal correlation is suppressed by a kinematic factor $Q
\sqrt{1-y}/\nu,$ where $y = {q \cdot p / p_e \cdot p},$ which
vanishes when the electron and photon are
parallel~\cite{Jaffe:2001dm,Barone:2001sp}.  Nevertheless, the
observed azimuthal spin asymmetry is large.  Large single spin
asymmetries have also been observed by the E704 experiment in $p p
\to \pi X$ reactions at Fermilab~\cite{Bravar:1996ki}.  In general,
such single-spin correlations require a correlation of a particle
spin with a production or scattering plane; they correspond to a
process which yields the $T-odd$ triple product $i \vec S_A \cdot
\vec p_B \times \vec p_C$ where the phase $i$ is required by
hermiticity and time-reversal invariance.  In the case of deeply
virtual Compton scattering, the phase can arise from the
interference of the real Bethe-Heitler and largely imaginary
Compton amplitude.  In purely hadronic processes, such correlations
can provide a window to the physics of final-state
interactions, an effect usually neglected in conventional QCD
analyses.

\eject

 In this brief theory summary talk of SPIN2001, I will give
an overview of recent progress in understanding proton spin and
spin correlations in quantum chromodynamics.  I will focus on
results which pose challenges to QCD, such as single-spin
asymmetries, the $A_{NN}$ asymmetry in large-angle elastic $pp$
scattering, the $J/\psi \to \rho \pi$ puzzle, and the problem of
$J/\psi$ polarization at the Tevatron.  An important tool is the
light-front wavefunction representation~\cite{PinskyPauli}, a
formalism in which the concepts of spin and orbital angular
momentum in relativistic bound states become transparent.  The
wavefunctions derived from light-front quantization play a central
role in QCD phenomenology, providing a frame-independent
description of hadrons in terms of their quark and gluon degrees
of freedom at the amplitude level~\cite{Brodsky:2001dx}.

\section{Relativistic Spin}

The spin decomposition of relativistic composite systems is
particularly transparent in light-front quantized QCD in light-cone gauge
$A^+ = 0,$ since all non-physical degrees of freedom of the quarks and
gluons are
eliminated by constraints~\cite{PinskyPauli,Srivastava:2000cf}.  Instead of
ordinary time
$t$, one quantizes the theory at fixed light-cone time \cite{Dirac:1949cp} $\tau
= t + z/c.$ The generator
$P^- = i {d\over d\tau}$ generates light-cone time translations, and
the eigen spectrum of the Lorentz scalar $ H^{QCD}_{LC} = P^+ P^-
- {\vec P_\perp}^2,$ where
$P^\pm = P^0 \pm P^z$ gives the mass spectrum of the color-singlet hadron
states,
together with their respective frame-independent light-front wavefunctions.
The total
spin projection $J_z$, as well as the momentum
generators
$P^+$ and
$\vec P_\perp,$ are kinematical; $i.e.$, they are independent
of the interactions.
For example, the
proton state satisfies:
$ H^{QCD}_{LC} \ket{\psi_p} = M^2_p \ket{\psi_p}$.  The expansion of
the proton eigensolution $\ket{\psi_p}$ on the color-singlet
$B = 1$, $Q = 1$ eigenstates $\{\ket{n} \}$
of the free Hamiltonian $ H^{QCD}_{LC}(g = 0)$ gives the
light-cone Fock expansion:
\begin{eqnarray}
\left\vert \psi_p(P^+, {\vec P_\perp} )\right> &=& \sum_{n}\
\prod_{i=1}^{n}
{{\rm d}x_i\, {\rm d}^2 {\vec k_{\perp i}}
\over \sqrt{x_i}\, 16\pi^3}\ \,
16\pi^3 \delta\left(1-\sum_{i=1}^{n} x_i\right)\,
\delta^{(2)}\left(\sum_{i=1}^{n} {\vec k_{\perp i}}\right)
\label{a318}
\\
&& \qquad \rule{0pt}{4.5ex}
\times \psi_n(x_i,{\vec k_{\perp i}},
\lambda_i) \left\vert n;\,
x_i P^+, x_i {\vec P_\perp} + {\vec k_{\perp i}}, \lambda_i\right>.
\nonumber
\end{eqnarray}
The light-cone momentum fractions $x_i = k^+_i/P^+$ and ${\vec
k_{\perp i}}$ represent the relative momentum coordinates of the
QCD constituents.  The physical transverse momenta are ${\vec
p_{\perp i}} = x_i {\vec P_\perp} + {\vec k_{\perp i}}.$ The
$\lambda_i$ label the light-front spin projections $S^z$ of the
quarks and gluons along the quantization direction
$z$~\cite{Lepage:1980fj}.  The corresponding spinors of the
light-front formalism automatically incorporate the Melosh-Wigner
transformation.  The physical gluon polarization vectors
$\epsilon^\mu(k,\ \lambda = \pm 1)$ are specified in light-cone
gauge by the conditions $k \cdot \epsilon = 0,\ \eta \cdot
\epsilon = \epsilon^+ = 0.$ The light-front wavefunctions
$\psi_{n/p}(x_i,\vec k_{\perp i},\lambda_i)$ of a hadron in QCD
are the projections of the hadronic eigenstate on the free
color-singlet Fock state $\ket n$ at a given light-cone time
$\tau.$ They thus represent the ensemble of quark and gluon states
possible when a hadron is intercepted at the light-front.  The
light-front wavefunctions are the natural extension of the
Schr\"odinger theory for many body theory; however, they are
Lorentz-invariant functions independent of the bound state's
physical momentum $P^+$, and $P_\perp$.  The parton degrees of
freedom are thus all physical; there are no Faddeev-Popov ghost or
other negative metric states~\cite{Srivastava:2000cf}.

The light-front formalism provides a remarkably transparent representation of
angular momentum~\cite{Ji:1996ek,%
Harindranath:1999ve,Brodsky:2001ii,%
Krassnigg:2001ka,Bashinsky:1998if,Jaffe:2001dm,PinskyPauli}.
The projection $J_z$ along the light-front quantization direction
is kinematical and is
conserved separately for each Fock component: each light-front Fock wavefunction
satisfies the sum rule:
$$ J^z = \sum^n_{i=1} S^z_i + \sum^{n-1}_{j=1} l^z_j
\ .  $$
The sum over $S^z_i$ represents the contribution of the
intrinsic spins of the $n$ Fock state constituents.  The sum over
orbital angular momenta
\begin{equation}
l^z_j = -{\mathrm i} \left(k^1_j\frac{\partial}{\partial k^2_j}
-k^2_j\frac{\partial}{\partial k^1_j}\right)
\end{equation}
derives from
the $n-1$ relative momenta and excludes the contribution to the
orbital angular momentum due to the motion of the center of mass,
which is not an intrinsic property of the
hadron~\cite{Brodsky:2001ii}.  The numerator structure of the
light-front wavefunctions in $k_\perp$ is thus largely determined by the
angular momentum constraints.
The spin properties of light-front wavefunctions
provide a consistent basis for analyzing the role of orbital angular
momentum and spin correlations and azimuthal spin
asymmetries in both exclusive and inclusive reactions.

The light-front representation thus provides a rigorous underpinning
for the familiar angular momentum sum rule for the
proton~\cite{Brodsky:1988ip,Jaffe:1989jz}:
$$J_z = {1\over 2} = {1\over 2} \Delta \Sigma + \Delta G + L_z.$$
The sum rule holds for each $n$ particle Fock state of the proton.
Here $\Delta \Sigma$ is the sum of the $S_z = \pm {1\over 2}$ of
the quarks and antiquarks.  The orbital term $L_z$ refers to the
sum over the $n-1$ internal orbital angular momenta of the quark
and gluon constituents.  In each case, the spin projections are
referenced to the light-front quantization direction $z$.  It is
easy to check that the QCD interactions of the light-cone
Hamiltonian conserve the total $J_z.$ Thus relativistic orbital
angular momentum has a physical, well-defined meaning in
light-cone gauge~\cite{Bashinsky:1998if}.

\section{Spin Dependent Local Matrix Elements}

The primary measures of deeply virtual inelastic lepton scattering
$\ell p \to \ell^\prime X$ and deeply virtual Compton scattering
$\gamma^* p \to \gamma p^\prime $ are the gauge invariant matrix
elements $\VEV{p^\prime | {\cal O} | p}$ of products of quark and
gluon fields at invariant separation $x^2 = {\cal O}({1\over
Q^2}),$ as generated by the operator product expansion.
Comprehensive discussions of the non-forward (or skewed) parton
distributions have been given by
Radyushkin~\cite{Radyushkin:1997ki} and Ji~\cite{Ji:1998pf}.
Polarization measurements are essential for distinguishing the various
operators.  For example, Ji~\cite{Ji:1998pf} has shown that there is a
remarkable connection of the $x$-moments of the chiral-conserving and
chiral-flip form factors $H(x,t,\zeta)$ and $ E(x,t,\zeta)$ which
appear in deeply virtual Compton scattering with the corresponding
spin-conserving and spin-flip electromagnetic form factors
$F_1(t)$ and $F_2(t)$ and gravitational form factors $A_{\rm
q}(t)$ and $B_{\rm q}(t)$ for each quark and anti-quark
constituent.  Thus, in effect, one can use virtual Compton
scattering to measure graviton couplings to the charged
constituents of a hadron~\cite{Diehl:2001kt}.

Given the light-front wavefunctions $\psi^{(\Lambda)}_{n/H},$ one
can construct any spacelike electromagnetic, electroweak, or
gravitational form factor or local operator product matrix element
from the diagonal overlap of the LC
wavefunctions~\cite{DY70,BD80}.  For example, the Pauli form factor
and the anomalous magnetic moment $\kappa = {e\over 2 M} F_2(0)$
are off-diagonal convolutions of light-front
wavefunctions~\cite{BD80}
\begin{equation}
-(q^1-{\mathrm i} q^2){F_2(q^2)\over 2M} =
\sum_a  \int
{{\mathrm d}^2 {\vec k}_{\perp} {\mathrm d} x \over 16 \pi^3}
\sum_j e_j \ \psi^{\uparrow *}_{a}(x_i,{\vec k}^\prime_{\perp
i},\lambda_i) \,
\psi^\downarrow_{a} (x_i, {\vec k}_{\perp i},\lambda_i)
{}\ ,
\label{LCmu}
\end{equation}
where the summation is over all contributing Fock states $a$ and
struck constituent charges $e_j$.  The arguments of the
final-state light-front wavefunction are $ {\vec k}'_{\perp
i}={\vec k}_{\perp i}+(1-x_i){\vec q}_{\perp}$ for the struck
constituent and $ {\vec k}'_{\perp i}={\vec k}_{\perp i}-x_i{\vec
q}_{\perp} $ for each spectator.  Notice that the anomalous magnetic moment
is calculated from the spin-flip non-forward matrix element
of the current where the incident and final wavefunctions have different
orbital angular momenta $\Delta L_z = \pm 1.$ In the
ultra-relativistic limit, where the radius of the system is small
compared to its Compton scale $1/M$, the anomalous magnetic moment
must vanish as required by the DHG sum rule~\cite{Bro94}.  The
light-cone formalism is consistent with this theorem.

The form factors of the energy-momentum tensor for a spin-\half \
system are defined by
\begin{eqnarray}
      \langle P'| T^{\mu\nu} (0)|P \rangle
       &=& \bar u(P')\, \Big[\, A(q^2)
       \gamma^{(\mu} \bar P^{\nu)} +
   B(q^2){i\over 2M} \bar P^{(\mu} \sigma^{\nu)\alpha}
q_\alpha \nonumber \\
   &&\qquad\qquad +  C(q^2){1\over M}(q^\mu q^\nu - g^{\mu\nu}q^2)
    \, \Big]\, u(P) \ ,
\label{Ji12}
\end{eqnarray}
where $q^\mu = (P'-P)^\mu$, $\bar P^\mu={1\over 2}(P'+P)^\mu$,
$a^{(\mu}b^{\nu)}={1\over 2}(a^\mu b^\nu +a^\nu b^\mu)$, and
$u(P)$ is the spinor of the system.  One can express the matrix
elements of the energy momentum tensor $T^{\mu \nu}$ as overlap
integrals of light-front wavefunctions~\cite{Brodsky:2001ii}.  As
in the light-cone decomposition of the Dirac and Pauli form
factors for the vector current, one can obtain the light-cone
representation of $A(q^2)$ and $B(q^2)$ from local matrix elements
of $T^{+ +}(0):$ An important consistency check of any
relativistic formalism is to verify the vanishing of the anomalous
gravito-magnetic moment $B(0)$, the spin-flip matrix element of
the graviton coupling and analog of the anomalous magnetic moment
$F_2(0)$.  For example, at one-loop order in QED, $B_f(0) =
{\alpha \over 3 \pi}$ for the electron when the graviton interacts
with the fermion line, and $B_\gamma(0) = -{\alpha \over 3 \pi}$
when the graviton interacts with the exchanged photon.  The
vanishing of $B(0)$ can be shown to be exact for bound or
elementary systems in the light-front
formalism~\cite{Brodsky:2001ii}, in agreement with the equivalence
principle~\cite{Okun,Kob70,Teryaev}.

In the case of the timelike matrix elements which appear in deeply
virtual Compton scattering and heavy hadron decays, one also has
contributions in which the initial and final Fock state number
differ by $\Delta n = 2.$ The skewed parton distributions which
control deeply virtual Compton scattering can also be defined from
diagonal and off-diagonal overlaps of light-front
wavefunctions~\cite{Brodsky:2000xy,Diehl:2000xz}.  The form factors
and matrix elements of local currents ${B|J^\mu(0)|A}$ which
appear in semileptonic decay amplitudes of heavy hadrons have an
exact representation as overlap momentum-space overlap integrals
of light-front wavefunctions~\cite{Brodsky:1999hn}.  The
light-front Fock representation is particularly important for
semi-leptonic exclusive matrix elements such as $B \to D \ell \bar
\nu$.  The Lorentz-invariant description requires both the overlap
of $n' = n$ parton-number conserving wavefunctions as well as the
overlap of wavefunctions with parton numbers $n' = n-2$ which
arises from the annihilation of a quark-antiquark pair in the
initial wavefunction~\cite{Brodsky:1999hn}.

The total angular
momentum projection of a state is given by~\cite{Ji:1998pf}
\begin{equation}
\VEV{J^i} = {1\over 2} \epsilon^{i j k} \int d^3x \VEV{T^{0 k}x^j - T^{0 j} x^k}
= A(0) \VEV{L^i} + \left[A(0) + B(0)\right] \bar u(P){1\over
2}\sigma^i u(P)
\ .
\label{Ji13a}
\end{equation}
The
$\VEV{L^i}$ term is the orbital angular momentum of the center of mass motion
with respect to an arbitrary origin and can be dropped.  The coefficient
of the $\VEV{L^i}$ term must be 1; $A(0) = 1 $ also follows when we evaluate
the four-momentum expectation value $\VEV{P^\mu}$.
Thus the total intrinsic angular momentum
$J^z$ of a nucleon can be identified with the values of the form factors
$A(q^2)$ and
$B(q^2)$ at
$q^2= 0:$
\begin{equation}
      \VEV{J^z} = \VEV{{1\over 2} \sigma^z} \left[A(0) + B(0)\right] \ .
\label{Ji13}
\end{equation}
One can define
individual quark and gluon contributions to the total
angular momentum from the matrix elements of the energy
momentum tensor \cite{Ji:1998pf}.
However, this definition is only formal; $A_{q,g}(0)$ can be interpreted
as the light-cone momentum fraction carried by the quarks or gluons
$\VEV{x_{q,g}}.$ The contributions from $ B_{q,g}(0) $ to $J_z$ cancel in the
sum.  In fact, as noted above,  the
contributions to $B(0)$ vanish when summed over the constituents of
each individual Fock state~\cite{Brodsky:2001ii}.

\section{The Spin Distributions of the Proton}

Given the light-front wavefunctions, one can compute the
$x$-dependence of the quark spin and
transversity distributions measured in polarized inclusive reactions.
For example,
the spin-polarized quark distributions at resolution (factorization scale)
$\Lambda$ correspond to~\cite{Lepage:1980fj,Brodsky:2001ii}
\begin{eqnarray}
q_{\lambda_q/\lambda_p}(x, \Lambda)
&=& \sum_{n,q_a}
\int\prod^n_{j=1} dx_j d^2 k_{\perp j}\sum_{\lambda_i}
\vert \psi^{(\Lambda)}_{n/H}(x_i,\vec k_{\perp i},\lambda_i)\vert^2
\\
&& \times \delta\left(1- \sum^n_i x_i\right) \delta^{(2)}
\left(\sum^n_i \vec k_{\perp i}\right)
\delta(x - x_q) \delta_{\lambda_a, \lambda_q}
\Theta(\Lambda^2 - {\cal M}^2_n)\ , \nonumber
\end{eqnarray}
where the sum is over all quarks $q_a$ which match the quantum
numbers, light-cone momentum fraction $x,$ and helicity of the
struck quark.  We thus have $g_1(x,\Lambda) = \Delta q(x,\Lambda)
= q(\lambda_q = +{1\over 2},\lambda_p = +{1\over 2}) - q(\lambda_q
= - {1\over 2},\lambda_p = +{1\over 2}).$ If the simple handbag
diagram of the parton model dominates the deep inelastic cross
section, then the quark distributions defined from the light-front
wavefunction probabilities will give an accurate representation of
the $g_1(x,Q)$ structure function in the Bjorken limit, where to
leading order in $\alpha_s$ we can identify the factorization
scale $Q$ and resolution scale $\Lambda.$ data.  These
distributions can also be defined as forward matrix elements of
gauge invariant local quark currents~\cite{Collins:1981uw}.

The most recent determinations of the integrated quark and gluon
distributions, as reported to this meeting by Ramsey,
are~\cite{Ramsey}
$$\Delta u = 0.86, \Delta d = - 0.40, \Delta s =
-0.06, \Sigma = \Delta q = 0.40, \Delta g = 0.44, L_z = - 0.1,$$
and $\Delta u_{sea} = -0.068.$ The error in each case is
approximately $\pm 0.04.$
 The effective factorization scale is
$Q^2 \simeq 1$ GeV$^2.$ The analysis incorporates recent E154
data, the Bjorken sum rule $\Delta u - \Delta d = g_A,$ and the
perturbative QCD ``counting rule" constraints $\Delta g(x) = x
g(x)$ and $\Delta q(x)/q(x) \to 1$ at $x \to 1$ which follow from
the dominance of the spin-aligned quark and gluon distributions.
This will be discussed in more detail in Section 5.  The results
show the presence of significant orbital angular momentum $L_z$
and gluon spin $\Delta g$, which is natural for light-cone
wavefunctions.  This will be illustrated in a QED example in
Section 9.  Applications to high $p_T$ inclusive processes in
polarized proton collisions are given by Gordon and
Ramsey~\cite{Gordon:1998hv}.

The gluon polarization also contributes to the quark and antiquark
spin distributions:  when the lepton scatters on a quark or
antiquark arising from gluon splitting, the quark and antiquark
contributions do not cancel.  The integration over quark
transverse momentum produces an anomalous
contribution~\cite{Altarelli:1988nr,Carlitz:ab,Anselmino:1994gn}
$-3 {\alpha_s\over 2\pi} \Delta G(x,Q)$ to the Ellis-Jaffe sum
rule~\cite{Ellis:1973kp} from quantum fluctuations.  When the
quark mass is high compared to the gluon virtuality, the anomaly
vanishes.  This decoupling follows from an application of the
Drell-Hearn Gerasimov sum rule~\cite{Bass:1998rn}.

Final-state interactions in gauge theory can affect deep inelastic
scattering reactions in a profound way~\cite{Brodsky:2001ue}.  The
rescattering of the outgoing quark leads to a leading-twist
contribution to the deep inelastic cross section from diffractive
channels $\gamma^* p \to q \bar q p^\prime,$ and the interference
effects induced by these diffractive channels cause nuclear
shadowing.  Nuclear shadowing, such as the strong spin-dependence
predicted for the deuteron spin structure
function~\cite{Melnitchouk:1994tx,Edelmann:1997ik,Piller:1999wx}
is thus not given by the nuclear light-cone wavefunctions.  The
final-state interactions affect the $x$ distributions at small
$x,$ and thus they can modify the extrapolations of the
spin-dependent structure functions into the low $x$ domain.
Similar complications may also affect the distributions obtained
from deeply virtual Compton scattering.  The moments of the
distributions which follow from the OPE are evidently not
affected.

There are in fact, possible phenomenological problems with the
perturbative QCD phenomenology.  Recent measurements of polarized
electron and proton deep inelastic scattering measurements by the
$E155$ collaboration at SLAC~\cite{Bosted:pd} give  \break
$\int^1_0 dx g^p_2(x,Q) = -0.034 \pm 0.008,$ in violation of the
Burkhardt-Cottingham sum rule~\cite{Burkhardt:ti}.  The second
moment of $g_2$ also appears to be inconsistent with model
predictions based on the Wandzura-Wilczek sum rule for
spin-$1\over 2$ partons~\cite{Wandzura:qf}.  Future plans to extend
these measurements at Jefferson laboratory were presented to this
meeting by J.-P. Chen and Z.-E.~Meziani.  V. Guzey also has
discussed complications due to the light nuclear targets used in
these experiments, including possible isobar
complications~\cite{Bissey:2001cw}.  Measurement of diffractive
dissociation $\ell A \to \ell^\prime p (A-1)$ can help to clarify
the complications of nuclear shadowing and final state
interactions~\cite{Brodsky:2001ue}.

When leading-twist factorization holds, one can predict spin
correlations for inclusive hadron production and jet measures
using the standard convolution of structure functions and
fragmentation functions, evolved by perturbative QCD.  An
excellent review of these phenomena is given by Boer and
Mulders~\cite{boermulders}.  Generalizations to next-to-leading
twist have been developed by Qiu and Sterman~\cite{Qiu:xx}.  The
analysis of azimuthal single-spin correlations of hadron spin with
the various production planes has been pioneered by
Collins~\cite{collins93,Barone:2001sp}.

\section{Dynamical Constraints on Polarized Structure functions and
Fragmentation Distributions}

One of the great challenges of nonperturbative QCD is to calculate
the magnitude and shapes of the quark and gluon spin distributions
from first principles in QCD.  In this meeting Melnitchouk reported
on lattice determinations of the distribution moments and the
complications of the chiral extrapolation~\cite{Detmold:2001dv}.
Another rigorous approach, the discretized light-front
quantization (DLCQ) method, computes the light-front wavefunctions
directly by diagonalizing the light-cone Hamiltonian on a discrete
Fock basis~\cite{Pauli:1985pv}.  Important progress in obtaining
explicit light-front wavefunctions for model 3+1 theories has been
reported by Hiller, McCartor, and myself~\cite{Brodsky:1999xj}.
There are also important guides obtained by QCD sum
rules~\cite{Ioffe:2000zd} and the meson cloud
model~\cite{Warr,Brodsky:1996hc,Cao:2001nu}.

Even without explicit nonperturbative solutions, one can use
theoretical constraints to pin down the large and small $x$
behavior of the spin distributions.  For example, one can use the
Regge behavior of high energy amplitudes to discern the power
behavior of spin-dependent fragmentation functions at small $x$ by
analyzing the spin or effective spin of particles coupling in the
$t$ channel in virtual Compton
scattering~\cite{Kuti:ph,Landshoff:1971ff,Brodsky:1973hm}.  On the
other hand, The behavior of structure functions where one quark
has the entire momentum requires the knowledge of LC wavefunctions
with $x \rightarrow 1$ for the struck quark and $x \rightarrow 0$
for the spectators.  This is a highly off-shell configuration, and
the fall-off the light-front wavefunctions at large $k_\perp$ and
$x \to 1$ is dictated by QCD perturbation theory since the state
is far-off the light-cone energy shell.  Thus one can rigorously
derive quark-counting and helicity-retention rules for the
power-law behavior of the polarized and unpolarized quark and
gluon distributions in the $x \rightarrow 1$ endpoint
domain~\cite{Lepage:1980fj,Brodsky:1994kg}.  Notice that $x\to 1$
corresponds to $k^z \to -\infty$ for any constituent with nonzero
mass or transverse momentum.

Modulo DGLAP evolution, the counting rule for the momentum
distribution for finding parton
$a$ in hadron $a$ at large $x \sim 1$ is $G_{a/A}(x) \propto (1-x)^{2
n_{\rm spect} - 1 + 2\vert \Delta S_z\vert}$ where $n_{\rm spect}$
is the minimum number of partons left behind when parton $a$ is
removed from $A$, and $\Delta S_z$ is the difference of the $a$
and $A$ spin projections.  This predicts $(1-x)^3$ behavior for valence
quarks with spin aligned with the parent proton's spin projection, and
$(1-x)^5$ behavior for anti-aligned quarks.  Similarly, the counting rules
predict that gluons with their spins aligned with the proton have
a $(1-x)^4$ distribution versus $(1-x)^6$ distribution when the
gluon and proton spins are anti-aligned.  These
constraints~\cite{Brodsky:1994kg} are usually incorporated into
phenomenological analyses such as the Ramsey \cite{Ramsey} and
MRST~\cite{Martin:2001es} approaches.  DGLAP evolution is quenched
in the large $x$ limit in the fixed $W^2$ domain One also expects
strong spin correlations in fragmentation functions.  Because of
the Gribov-Lipatov relation, the same counting rule behavior holds
for fragmentation $D_{H \to a}(z,Q)$ distributions at $z \to 1.$
This is particularly interesting in the case of the fragmentation
of a quark into a $\Lambda$ since its polarization is
self-analyzing.

In the large $x$ or $z$ domain, higher twist terms involving
several partons of the target can dominate inclusive rates since
fewer spectators are stopped~\cite{Berger:1979du}.  The clearest
example is the Drell-Yan process $\pi A \to \ell^+ \ell^- X$ at
$x_1 \to 1$ where the process $\pi q \to \gamma^* q$ subprocess
can dominate the leading twist $\bar q(x_1) q(x_2) \to \gamma^*
\to \ell^+ \ell^-$ amplitude.  In the higher twist process the
virtual photon is produced with longitudinal polarization.  The
muon differential cross section thus has the form
\begin{equation}{d\sigma\over d \cos \theta d x_1} = A(1-x_1)^2(1+
\cos^2\theta) + {B\over Q^2} \sin^2 \theta,
\end{equation}
 in good
agreement with the data.  Thus at large momentum fraction $x_1,$
the higher-twist term is amplified by a factor ${\mu^2 /[ Q^2
(1-x_1)^2]}.$ In addition, the coplanarity distribution of the muon
pair plane with the $\pi \to \gamma^*$ production plane is
consistent with these expectations~\cite{McGaughey:1999mq}.

\section{Proton Spin Carried by Non-Valence Quarks}

Since particle number is not conserved in a relativistic theory,
the wavefunctions of hadrons must contain non-valence states,
corresponding to additional sea quarks and gluons in flight at a
fixed light-cone time $\tau$; part of this fluctuation corresponds
to the resolved substructure of the valence quarks and is
associated with DGLAP evolution.  However, the sea quarks and
gluons which are multiply connected to the valence quarks are
intrinsic to the hadron's structure and are not generated by
perturbative evolution~\cite{Brodsky:1980pb}.  One can show
rigorously from the OPE, that the momentum and spin carried by the
intrinsic sea quarks in non-Abelian QCD decreases as $1/m_Q^2$ for
heavy quarks like charm~\cite{Franz:2000ee}.  If one associates
the intrinsic strange quarks in the proton with its $K^+ \Lambda$
fluctuation, then the strange quark spin will be anti-aligned with
the nucleon and the anti-strange quark will be unpolarized since
it is associated with the pseudoscalar
meson~\cite{Brodsky:1996hc}.  The momentum fraction carried by
intrinsic charm quarks is of the order of
$1\%$\cite{Harris:1995jx}, and thus one expects the fraction of
proton spin carried by carried by charm quarks to also be of this
order~\cite{Polyakov:1998rb,Song:2001kh}.

\section{Heavy Quarkonium Polarization in Inclusive Reactions}

There are still many unresolved problems involving of $J/\psi$
production in high energy hadron collisions.  The expectation from
the color-singlet model that the $J/\psi$ will be produced with
transverse polarization at large transverse momentum in $p p$
collisions has not been confirmed~\cite{Mizukoshi:1999ja}.  In the
case of $\pi A \to J/\psi X$, the $J/\psi$ is generally produced
unpolarized.  However at $x_F \sim 0.95$ the $J/\psi$ is observed
to have significant longitudinal polarization~\cite{Biino:1987qu}.
This anomalous behavior can be understood if the dominant
subprocess is the diffractive excitation of the $q \bar q c \bar
c$ Fock state of the pion at large $x_F, $ since the $c \bar c$
pair will have the same in as the parent pion when they have a
large momentum fraction $x_F = x_c + x_{\bar c} \to
1$~\cite{Brodsky:1991dj}.

\section{Transversity}

The transversity distribution $\delta q(x,Q)$
gives the correlation between transversely polarized
quarks and the transverse polarization of the parent
hadron~\cite{Ralston:ys,Collins:1993kq,Jaffe:1993xb}.
Transversity is a leading-twist distribution which
can be measured in inclusive processes with two hadrons such as the Drell-Yan
process.  An excellent review of the theory and phenomenology of transversity
was presented to this meeting by Jaffe~\cite{Jaffe:2001dm}.

Transversity can be determined from the light-front wavefunctions
of the target.  A transversely polarized spin-\half particle
polarized in the $\hat y$ direction can be represented as the
state $\ket N = {|+> - |->\over \sqrt 2}$ where $\ket \pm$
represents the $S_z = \pm {1\over 2}$ state.  Thus the
transversity distribution is a density matrix of light-front
wavefunctions:
\begin{eqnarray}
\delta q(x, \Lambda) &=& \sum_{n,q_a} \int\prod^n_{j=1} dx_j d^2
k_{\perp j}\sum_{\lambda_i}\Big|
 \psi^{(\Lambda) *}_{n/H}(x_i,\vec k_{\perp i},\lambda_i=\lambda_p=+{1\over2})
\\
&&\times \qquad \qquad \qquad \qquad \qquad
 \psi^{(\Lambda)}_{n/H}(x_i,\vec k_{\perp
i},\lambda_i=\lambda_p=-{1\over2})\Big|
\\
&& \times ~ \delta\left(1- \sum^n_i x_i\right) \delta^{(2)}
\left(\sum^n_i \vec k_{\perp i}\right)
\delta(x - x_q) \delta_{\lambda_a, \lambda_q}
\Theta(\Lambda^2 - {\cal M}^2_n)\ , \nonumber
\end{eqnarray}
The Soffer inequality~\cite{Soffer:1994ww} $\delta q(x,\Lambda) \le {1\over
2}[\Delta q(x,\Lambda)+ q(x,\Lambda)]$ follows simply from the light-cone
representation.  The integrated
transversity is the matrix element of the quark tensor charge
$\bar \psi(0) \sigma^{+i}\gamma^5 \psi(0)$ operator with quantum numbers
$J^{CP}= 1^{+-}.$ Gronberg and Goldstein~\cite{Gamberg:2001qc}
have shown how one can use the measured
couplings of axial vector meson to the proton in order to estimate the
strength of the transversity of the $u$ and $d$ quarks in the proton at a
factorization scale $x^2 = 1/\mu^2$ of the order of the size of hadrons.
As a simple example of transversity, the spin decomposition of the
electron in QED due to its lowest order quantum fluctuations will be analysed in
the next section.

\section{The Light-Front Wavefunctions and Spin Structure of
Leptons in QED}

The light-front wavefunctions of a lepton in QED provide an ideal
system to check explicitly the intricacies of relativistic spin
and angular momentum in quantum field
theory~\cite{Brodsky:2001ii}.  Although they are derived in
perturbation theory,  the light-front wavefunctions of leptons and
photons can be used as templates for the wavefunctions of
non-perturbative composite systems resembling hadrons in QCD.  For
generality, one can assign a mass $M$ to the external lepton, a
different mass $m$ to the internal fermion lines, and a mass
$\lambda$ to the internal photon line, thus simulating the
quark-spin-$1$ diquark structure of a baryon.  The fast-falling
wavefunction of a composite state can be simulated by
differentiating the wavefunction with respect to a mass parameter
or employing a Pauli-Villars spectrum.

The two-particle Fock state for an electron with $J^z = + {1\over
2}$ has
\begin{equation}
\left
\{ \begin{array}{l}
\psi^{\uparrow}_{+\frac{1}{2}\, +1} (x,{\vec k}_{\perp})=-{\sqrt{2}}
\frac{(-k^1+{\mathrm i} k^2)}{x(1-x)}\,
\varphi \ ,\\
\psi^{\uparrow}_{+\frac{1}{2}\, -1} (x,{\vec k}_{\perp})=-{\sqrt{2}}
\frac{(+k^1+{\mathrm i} k^2)}{1-x }\,
\varphi \ ,\\
\psi^{\uparrow}_{-\frac{1}{2}\, +1} (x,{\vec k}_{\perp})=-{\sqrt{2}}
(M-{m\over x})\,
\varphi \ ,\\
\psi^{\uparrow}_{-\frac{1}{2}\, -1} (x,{\vec k}_{\perp})=0\ ,
\end{array}
\right.
\label{vsn2}
\end{equation}
where
\begin{equation}
\varphi=\varphi (x,{\vec k}_{\perp})=\frac{ e/\sqrt{1-x}}{M^2-({\vec
k}_{\perp}^2+m^2)/x-({\vec k}_{\perp}^2+\lambda^2)/(1-x)}\ ,
\label{wfdenom}
\end{equation}
with similar expressions for $\psi^{\downarrow}.$ The coefficients
of $\varphi$ are the matrix elements of
$\frac{\overline{u}(k^+,k^-,{\vec k}_{\perp})}{{\sqrt{k^+}}}
\gamma \cdot \epsilon^{*} \frac{u (P^+,P^-,{\vec
P}_{\perp})}{{\sqrt{P^+}}}$ which are the numerators of the
wavefunctions corresponding to each constituent spin $s^z$
configuration.  The two boson polarization vectors in light-cone
gauge are $\epsilon^\mu= (\epsilon^+ = 0\ , \epsilon^- = {\vec
\epsilon_\perp \cdot \vec k_\perp \over 2 k^+}, \vec
\epsilon_\perp)$ where
$\vec{\epsilon}=\vec{\epsilon_\perp}_{\uparrow,\downarrow}= \mp
(1/\sqrt{2})(\widehat{x} \pm {\mathrm i} \widehat{y})$.  The
polarizations also satisfy the Lorentz condition $ k \cdot
\epsilon =0$.

The transverse momentum dependence of the numerator of each
light-front wavefunction specifies the orbital angular momentum
guarantees $J_z$ conservation for each Fock state.  The fermion
spin is given by the matrix element of the light-cone spin
operator $\gamma^+\gamma^5$ \cite{Ma91b}; the relative orbital
angular momentum operator is $-{\mathrm i}
(k^1\frac{\partial}{\partial k^2} -k^2\frac{\partial}{\partial
k^1})$ \cite{Harindranath:1998ve,MS98,Hag98,Song:nw}.  Each
configuration satisfies the spin sum rule: $J^z=s^z_{\rm
f}+s^z_{\rm b} + l^z$.

In the non-relativistic limit, the transverse motion of the
constituents can be neglected and we have only the
$\ket{+\frac{1}{2}} \to \ket{-\frac{1}{2}\, +1}$ configuration
which is the non-relativistic quantum state for the spin-half
system composed of a fermion and a spin-1 boson constituents.  The
fermion constituent has spin projection in the opposite direction
to the spin $J^z$ of the whole system.  However, for
ultra-relativistic binding in which the transverse motions of the
constituents are large compared to the fermion masses (and at large
evolution scales),  the
$\ket{+\frac{1}{2}} \to \ket{+\frac{1}{2}\, +1}$ and
$\ket{+\frac{1}{2}} \to \ket{+\frac{1}{2}\, -1}$ configurations
dominate over the $\ket{+\frac{1}{2}} \to \ket{-\frac{1}{2}\, +1}$
configuration.  In this case the fermion constituent has spin
projection parallel to $J^z$.  In the case of Yukawa theory
corresponding to a spin-0 diquark, the non-relativistic fermion's
spin projection is aligned with the total $J^z$, and it is
anti-aligned in the ultra-relativistic limit.

Given the light-front wavefunctions, one can explicitly calculate
the helicity-flip electromagnetic and gravitational form factors
for the fluctuations of the electron at one-loop, and verify the
Schwinger anomalous moment and the cancellation of the sum of
graviton couplings $B(q^2)$ to the constituents at $q^2 =
0$~\cite{Brodsky:2001ii}.  The contribution to $B(q^2)$ where the
photon couples to the photon constituent has threshold dependence
$\sim \sqrt{-q^2/ m^2}$ reflecting the massless two-body cut in
the $t$ channel.

The spin distributions in the QED model are also
easily computed:
\begin{eqnarray}
&&\Delta q(x,\Lambda^2)_{\rm spin-1\ diquark}
\label{dqvector} \\
&=&
\int\frac{{\mathrm d}^2 {\vec k}_{\perp} {\mathrm d} x }{16 \pi^3}
\theta (\Lambda^2 - {\cal M}^2)
\ 2\ \Big[ \ {{\vec k}_{\perp}^2\over x^2(1-x)^2}\
+\ {{\vec k}_{\perp}^2\over (1-x)^2}\
-\ (M-{m\over x})^2\ \Big]\ |\varphi |^2\ .
\nonumber
\end{eqnarray}
In the case of the Yukawa model, where the boson plays the role of a spin-0
diquark, one finds
\begin{equation}
\Delta q(x,\Lambda^2)_{\rm spin-0\ diquark}=
\int\frac{{\mathrm d}^2 {\vec k}_{\perp} {\mathrm d} x }{16 \pi^3}
\theta (\Lambda^2 - {\cal M}^2)
\ \Big[ \
(M+{m\over x})^2\ -\ {{\vec k}_{\perp}^2\over x^2}\ \Big]\ |\varphi |^2\ ,
\label{dqscalar}
\end{equation}
where we have regulated the integral by assuming a cutoff in the
invariant mass: ${\cal M}^2 = \sum_i {{\vec k}^2_{\perp i}
+m^2_i\over x_i} < \Lambda^2.$ In the spin-0 diquark model $\Delta
q = 1$ in the nonrelativistic limit, and decreases toward $\Delta
q = - 1$ as the intrinsic transverse momentum increases.  The
behavior is just opposite in the case of the spin-1 diquark.  The
distinct features of spin structure in the non-relativistic and
ultra-relativistic limits reveals the importance of relativistic
effects and supports the viewpoint \cite{Ma91b,Ma96,MSS97} that
the proton ``spin puzzle" can be understood as due to the
relativistic motion of quarks inside the nucleon.  In particular,
the spin projection of the relativistic constituent quark tends to
be anti-aligned with the proton spin in a quark-diquark bound
state if the diquark has spin 0.  The state with orbital angular
momentum $l^z= \pm 1 $ in fact dominates over the states with $l^z
= 0.$ Thus the empirical fact that $\Delta q$ is does not saturate
the spin of the proton has a natural description in the light-cone
Fock representation of hadrons.

In the case of transversity, we require the overlap of the
$\psi^{* \uparrow}_{+{1\over 2}}$ and $\psi^\downarrow_{-{1\over 2}}$ states.
Thus
\begin{eqnarray}
\delta q(x,\Lambda^2)_{\rm spin-1\ diquark}
\label{dqvector1}
=
\int\frac{{\mathrm d}^2 {\vec k}_{\perp} {\mathrm d} x }{16 \pi^3}
\theta (\Lambda^2 - {\cal M}^2)
\ 2\ \Big[ \ {{\vec k}_{\perp}^2\over x(1-x)^2}\
 \Big]\ |\varphi |^2\ .
\nonumber
\end{eqnarray}
In the case of spin-0 diquarks, one finds
\begin{equation}
\delta q(x,\Lambda^2)_{\rm spin-0\ diquark}=
\int\frac{{\mathrm d}^2 {\vec k}_{\perp} {\mathrm d} x }{16 \pi^3}
\theta (\Lambda^2 - {\cal M}^2)
\ \Big[ \
(M+{m\over x})^2\ \Big]\ |\varphi |^2\ ,
\label{dqscalar1}
\end{equation}
In each case the result obeys the Soffer inequality~\cite{Soffer:1994ww}.

\section{Spin and Exclusive Processes}

One of the important new areas of spin phenomenology is the exclusive
decays of the $B$ mesons~\cite{Beneke:2000ry,Keum:2000wi,Brodsky:2001jw}.  The
factorization theorems for hard exclusive processes~\cite{Brodsky:1989pv} in
which amplitudes factorize as products of distribution amplitudes and hard
scattering quark and gluon scattering amplitudes have been generalized to
exclusive $B$ decays.  The spin observables provide sensitive tests of
factorization and the prediction of diminished final-state interactions (color
transparency)~\cite{Brodsky:xz}.  Predictions for $B \to J/\psi K^*$ and
$\Lambda_B$ decay were presented at this meeting by K.-C.
Yang~\cite{Cheng:2001cs}.  The physics of the $B$ distribution amplitude using
constraints from heavy quark symmetry was presented by
Kawamura~\cite{Kawamura:2001bp}.
Intrinsic charm in the $B$ wavefunction can
play an enhanced role in its weak decays because of the CKM
hierarchy~\cite{Chang:2001yf,Brodsky:2001yt}.

The features of exclusive processes to leading power in the transferred
momenta are well known~\cite{Brodsky:1989pv}:

(1) The leading power fall-off is given by dimensional counting
rules for the hard-scattering amplitude: $T_H \sim 1/Q^{n-1}$,
where $n$ is the total number of fields (quarks, leptons, or gauge
fields) participating in the hard scattering
\cite{BF,Matveev:1973ra}.  Thus the reaction is dominated by
subprocesses and Fock states involving the minimum number of
interacting fields.  The hadronic amplitude follows this fall-off,
modulo logarithmic corrections from the running of the QCD
coupling and the evolution of the hadron distribution amplitudes.
In some cases, such as large angle $p p \to p p $ scattering,
pinch contributions from multiple hard-scattering processes must
also be included \cite{Landshoff:1974ew}.  The general success of
dimensional counting rules implies that the effective coupling
$\alpha_V(Q^*)$ controlling the gluon exchange propagators in
$T_H$ are frozen in the infrared, \ie, have an infrared fixed
point, since the effective momentum transfers $Q^*$ exchanged by
the gluons are often a small fraction of the overall momentum
transfer \cite{Brodsky:1998dh}.  The pinch contributions are then
suppressed by a factor decreasing faster than a fixed power
\cite{BF}.

(2) The leading power dependence is given by hard-scattering
amplitudes $T_H$ which conserve quark helicity
\cite{Brodsky:1981kj,Chernyak:1999cj}.  Since the convolution of
$T_H$ with the light-front wavefunctions projects out states with
$L_z=0$, the leading hadron amplitudes conserve the sum of
light-cone spin projections $s^z_i.$ Thus the sum of initial and
final hadron helicities are conserved.  Hadron helicity
conservation thus follows from the underlying chiral structure of
QCD.

One of the important predictions of hadron helicity conservation
in perturbative QCD is the relative suppression of the Pauli
versus the Dirac form factors: ${F_2(Q^2)/ F_1(Q^2) }\to {\mu^2/
Q^2}$ (modulo logarithmic factors) for timelike and spacelike form
factors at high $Q^2$.  However recent
measurements~\cite{Jones:1999rz} at Jefferson laboratory using the
ratio of transverse and longitudinal recoil proton polarization
correlated with the electron longitudinal spin gives the ratio
$G^p_E(Q^2)/ G^p_M(Q^2)$ directly; the results are consistent with
a behavior ${F_2(Q^2)/ F_1(Q^2)} \sim {C / Q}$ for $2 < Q^2 < 5$
GeV$^2.$ The experimental results suggest that the Pauli form
factor may be receiving significant logarithmic enhancement
factors, possibly from the Sudakov-controlled fixed $k_\perp,$ $x
\to 1$ integration regime.  It is also important to check the
angular distribution in $p \bar p \to \ell^+ \ell^-$ and $e^+ e^-$
annihilation to baryon pairs.  Hadron helicity conservation
predicts the dominance of $1+\cos^2 \theta$ distributions at large
$s.$

Hadron-helicity conservation also predicts the suppression of
vector meson states produced with $J_z =\pm 1$ in $e^+ e^=$
annihilation to vector-pseudoscalar final
states~\cite{Brodsky:1981kj}.  However, $J/\psi \to \rho \pi$
appears to occur copiously whereas $\psi^\prime \to \rho \pi$ has
never been conserved.  The PQCD analysis assumes that a heavy
quarkonium state such as the $J/\psi$ always decays to light
hadrons via the annihilation of its heavy quark constituents to
gluons.  However, as Karliner and I~\cite{Brodsky:1997fj} have
shown, the transition $J/\psi \to \rho \pi$ can also occur by the
rearrangement of the $c \bar c$ from the $J/\psi$ into the $\ket{
q \bar q c \bar c}$ intrinsic charm Fock state of the $\rho$ or
$\pi$.  On the other hand, the overlap rearrangement integral in
the decay $\psi^\prime \to \rho \pi$ will be suppressed since the
intrinsic charm Fock state radial wavefunction of the light
hadrons will evidently not have nodes in its radial wavefunction.
This observation can provide a natural explanation of the
long-standing puzzle why the $J/\psi$ decays prominently to
two-body pseudoscalar-vector final states, whereas the
$\psi^\prime$ does not.

One of the most striking anomalies in elastic proton-proton
scattering is the large spin correlation $A_{NN}$ observed at
large angles \cite{krisch92}.  At $\sqrt s \simeq 5 $ GeV, the rate
for scattering with incident proton spins parallel and normal to
the scattering plane is four times larger than that for scattering
with anti-parallel polarization.  This strong polarization
correlation can be attributed to the onset of charm production in
the intermediate state at this energy
\cite{Brodsky:1988xw,deTeramond:1998ny}.  A resonant intermediate
state $\vert u u d u u d c \bar c \rangle$ has odd intrinsic
parity and can thus couple to the $J= L=S=1$ initial state, thus
strongly enhancing scattering when the incident projectile and
target protons have their spins parallel and normal to the
scattering plane.  The charm threshold can also explain the
anomalous change in color transparency observed at the same energy
in quasi-elastic $ p p$ scattering.  A crucial test is the
observation of open charm production near threshold with a cross
section of order of $1 \mu$b.  Analogous strong spin effects
should also appear at the strangeness threshold and in exclusive
photon-proton reactions such as large angle Compton scattering and
pion photoproduction near the strangeness and charm thresholds.  An
alternate hypothesis, based on the interference of Landshoff pinch
contributions has been proposed by Jain and
Ralston~\cite{Jain:2000rg}.

An important new area of study is exclusive channels in electroproduction,
especially for longitudinally polarized virtual photons, where factorization
theorems can also be proved~\cite{Collins:1999yw}.

The application of perturbative QCD to exclusive nuclear processes
such as the deuteron form
factors,~\cite{Brodsky:1976rz,Brodsky:1983vf,Brodsky:1992px}
photodisintegration~\cite{Brodsky:1983kb}, and meson
photoproduction~\cite{Brodsky:2001qm} on a polarized deuteron
target is particularly interesting as first principle tests of the
applicability of QCD to nuclear physics.  For example, the
dominance of helicity-conserving amplitudes in gauge theory can be
shown to imply universal ratios for the charge, magnetic, and
quadrupole form factors of spin-one bound
state~\cite{Brodsky:1992px}.  These results provide all-angle
predictions for the leading power behavior of the tensor
polarization $T_{20}(Q^2,\theta_{cm})$ and the invariant ratio
$B(Q^2)/A(Q^2),$ although significant higher-twist corrections are
expected at presently accessible momentum transfers.  One can
also show that the magnetic and quadrupole moments of any composite
spin-one system take on the canonical values $\mu=e/M$ and $Q=-e/M^2$ in
the strong binding limit of the zero bound-state radius or infinite
excitation energy independent of the internal dynamics.  A comprehensive
review of the status of spin tests in exclusive nuclear processes and the
successes and failures of the perturbative QCD approach is given
by Gilman and Gross~\cite{Gilman:2001yh}.  It should be emphasized that
the normalization of the perturbation theory deuteron matrix elements are
strongly affected by the existence of six-quark hidden color Fock
states~\cite{Brodsky:1983vf}.

\section{Single-Spin Asymmetries}

Single-spin asymmetries in hadronic reactions have been among the most
difficult phenomena to understand from basic principles in QCD.  The
problem has become more acute because of the observation
by the HERMES~\cite{hermes0001} and SMC~\cite{smc99} collaborations of a strong
correlation between the target proton spin $\vec S_p$ and the plane of a
produced pion in semi-inclusive deep inelastic lepton scattering $\ell
p^\uparrow \to \ell^\prime \pi X$ at photon virtuality as large as $Q^2=
6$ GeV$^2$.  Large azimuthal single-spin asymmetries have also been seen
in hadronic reactions such as $p p^\uparrow \to \pi X$ \cite{Bravar:1996ki},
where the target antiproton is polarized normal to the pion production
plane, and in $ p p \to \Lambda^\uparrow X$ ~\cite{Skeens:1991my}, where the
hyperon is polarized normal to the $\Lambda$ production plane.

In the target rest frame, single-spin correlations correspond to
the $T-odd$ triple product $i \vec S_p \cdot \vec p_\pi \times
\vec q,$ where the phase $i$ is required by time-reversal
invariance.  In order to produce such an azimuthal correlation involving
a transversely polarized proton, there are two necessary conditions:
(1) There must be two proton spin amplitudes $M[{\gamma^*
p(J^z_p)\to F}]$ with $J^z_p = \pm {1\over 2}$ which couple to the
same final-state $\ket F$; and (2) The two amplitudes must have
different, complex phases.  The analysis of single-spin
asymmetries thus requires an understanding of QCD at the amplitude
level, well beyond the standard treatment of hard inclusive
reactions based on the factorization of structure functions and
fragmentation functions.  Since we need the interference of two
amplitudes which have different proton spin $J^z_p= \pm {1\over
2}$ but couple to the same final-state, the orbital angular
momentum of the two proton wavefunctions must differ by $\Delta
L^z = 1.$ If a target is stable, its light-front wavefunction must
be real.  Thus the only source of a nonzero complex phase in
leptoproduction in the light-front frame are final-state
interactions.  The rescattering corrections from final-state
exchange of gauge particles produce Coulomb-like complex phases
which, however, depend on the proton spin.  Thus $M[{\gamma^*
p(J^z_p= \pm {1\over 2} ) \to F}]= |M[{\gamma^* p(J^z_p=\pm
{1\over 2} ) \to F}]|\, e^{i \chi_\pm}$.  Each of the phases is
infrared divergent; however the difference $\Delta \chi = \chi_+ -
\chi_-$ is infrared finite and nonzero.  The resulting single-spin
asymmetry is then proportional to ${\rm sin} \Delta \chi.$

Recently, Dae Sung Hwang, Ivan Schmidt and I~\cite{BHS} have found
that final-state interactions from gluon exchange between the
outgoing quark and the target spectator system leads to
single-spin asymmetries in deep inelastic lepton-proton scattering
at leading twist in perturbative QCD; {\em i.e.}, the rescattering
corrections are not power-law suppressed at large photon
virtuality $q^2$ at fixed $x_{bj}$.  The existence of such
single-spin asymmetries requires a phase difference between two
amplitudes coupling the proton target with $J^z_p = \pm {1\over
2}$ to the same final-state, the same amplitudes which are
necessary to produce a nonzero proton anomalous magnetic moment.
We find that the exchange of gauge particles between the outgoing
quark and the proton spectators produces a Coulomb-like complex
phase which depends on the angular momentum $L^z$ of the proton's
constituents and thus is distinct for different proton spin
amplitudes.  The single-spin asymmetry which arises from such
final-state interactions does not factorize into a product of
structure function and fragmentation function, and it is not
related to the transversity distribution $\delta q(x,Q)$ which
correlates transversely polarized quarks with the spin of the
transversely polarized target nucleon.

We have calculated the single-spin asymmetry in semi-inclusive
electroproduction $\gamma^* p \to H X$ induced by final-state
interactions in a model of a spin-\half ~ proton of mass $M$
composed of charged spin-\half - spin-0 constituents of mass $m$
and $\lambda$, respectively, as in the QCD-motivated quark-diquark
model of a nucleon.  The azimuthal single-spin asymmetry ${\cal
P}_y$ transverse to the photon-to-pion production plane decreases
as $ \alpha_s(r^2_\perp) x_{bj} M r_\perp [\ln r^2_\perp]/ {\vec
r}_{\perp}^2$ for large $r_\perp,$ where $r_\perp$ is the
magnitude of the momentum of the current quark jet relative to the
$q$ direction.  The mass $M$ of the physical proton mass appears
here since it determines the ratio of the $L_z = 1$ and $L_z = 0$
matrix elements.  The final-state interactions from gluon exchange
between the outgoing quark and the target spectator system leads
to single-spin asymmetries in deep inelastic lepton-proton
scattering at leading twist in perturbative QCD; {\em i.e.}, the
rescattering corrections are not power-law suppressed at large
photon virtuality $q^2$ at fixed $x_{bj}$.  The linear fall-off in
$r_\perp$ compensates for the higher twist of the $q \bar q$ gluon
correlation.  The nominal size of the spin asymmetry is thus $C_F
\alpha_s(r^2_\perp) a_p$ where $a_p$ is the proton anomalous
magnetic moment.  Our analysis shows that the single-spin asymmetry
which arises from final-state interactions does not factorize
since the result depends on the $\VEV{p |\bar \psi_q A \psi |p}$
proton correlator, not the usual quark distribution derived from
$\VEV{p |\bar \psi_q (\xi) \psi_q(0)|p}$ evaluated at equal
light-cone time $\xi^+ = 0$.  We thus predict that the single-spin
asymmetry in electroproduction is independent of $Q^2$ at fixed
$\Delta = x_{bj}$.  This approach also can be applied to
single-spin asymmetries in more general hadronic hard inclusive
reactions such as $p p \to \Lambda X.$

In the case of deeply virtual Compton scattering $\ell p \to
\ell^\prime \gamma(k) X$, the HERMES
collaboration~\cite{Airapetian:2001yk} has measured a large single
spin asymmetry corresponding to the azimuthal correlation $i\vec
S_p \cdot \vec k \times \vec q.$ The phase of the Bethe-Heitler
amplitude is real, whereas the complex phase of the virtual
Compton amplitude~\cite{Brodsky:1973hm} is dictated by Regge exchange in
the $t-$channel: $M_{\rm Compt}\propto {(s/Q^2)}^{\alpha_R(t) }[1 + e^{i
\pi \alpha_R(t)}].$ This is one source of phase interference.  In
addition, the same final state interaction which provides a phase
in $\gamma^* p \to \pi p$ will also contribute to the deeply
virtual Compton amplitude.

Recently HERMES has reported a significant single-spin azimuthal asymmetry in
exclusive electroproduction of $\pi^+$ mesons~\cite{Airapetian:2001iy}.  These
new measurements will open up an important new window to the effects of
rescattering in hard exclusive reactions~\cite{Collins:1999yw} Studies of
single-spin asymmetries are also of critical importance in
$B$ and $B^*$ decays since the presence of final-state hadronic
interactions have to be understood in order to interpret CP-violating
parameters~\cite{Bigi:1997fj}.

\section{Acknowledgements}
I thank Professor Bo-Qiang Ma and his colleagues at Peking
University and the Institute for High Energy Physics for
organizing this exceptional meeting and for their outstanding
hospitality in Beijing.  I also thank P. Bosted, M. Burkardt,
M.~Diehl, L. Gamberg, R. Jaffe, J. Hiller, D. S. Hwang, X. Ji, Y.
Y. Keum, H. N. Li, G. Ramsey, I.~Schmidt, and J. Soffer, as well
as many of the other participants of this meeting for helpful
discussions.

\end{document}